\documentclass[
    ,final            
  ]
  {aipproc}

\layoutstyle{6x9}
\newcommand\apjl{Ap. J. Lett.}
\newcommand\apjs{Ap. J. Supp.}
\newcommand\apj{Ap. J.}
\newcommand\mnras{Mon. Not. RAS}
\newcommand\nat{Nature}
\newcommand\physrep{Phys. Rep.}

\newcommand\aap{Astron. \& Astrophys.}

\begin{document}

\title{Magnetic Fields in Gamma-Ray Bursts: A Short Overview}


\author{Tsvi Piran}{
  address={Racah Institute for Physics, The Hebrew University, Jerusalem, 91904
  Israel\\
  Theoretical Astrophysics, Caltech, Pasadena, CA 91125, USA}
}

\begin{abstract}
Magnetic fields play a crucial role in the physics of Gamma-Ray
Bursts (GRBs). Strong observational evidence indicates that the
observed afterglow and most likely the prompt emission arise from
synchrotron emission.  It is possible that  Poynting flux plays an
important or even dominant role in the relativistic outflow from
the inner engine, but like in other astronomical relativistic jets
this suggestion is controversial. Finally, it is likely that
magnetic fields larger than $10^{15}$ G occur within GRBs' inner
engines and contribute to the  acceleration and collimation of the
relativistic jets. I review here the GRB fireball model and
discuss the role that magnetic fields play in its various
components. I suggest that the early afterglow, that reflects the
initial interaction of the relativistic jet with its surrounding
matter is the best available tool to explore the nature of
relativistic outflow in astronomical relativistic jets.
\end{abstract}

\maketitle


\section{Introduction}

More than thirty years, after the discovery of Gamma-Ray bursts
(GRBs) we have now a reasonable GRB model. The model is based on
the dissipation of an ultra-relativistic outflow. At first the
flow is dissipated by internal shocks (or another form of internal
dissipation) that produce the prompt $\gamma$-rays. Later the
interaction of the flow with the circum-burst matter produces an
external shock and this blast wave produces the subsequent
afterglow.

The role of magnetic fields varies within GRB models from crucial
to ultimate. It is generally accepted that the observed afterglow
is produced by synchrotron emission
\cite{Katz94s,MesReesWei97,SPN98,lp00}. Synchrotron is also the
best-bet model for the prompt $\gamma$-rays emission (note however
some criticism of the synchrotron model for the prompt emission
\cite{CohenKatzP98,Preece02,sp04}). Even within the  conventional
model of a baryonic relativistic outflow \cite{SP90,p90,MR92}
magnetic fields within the shocked regions are essential to
produce this synchrotron emission. In the maximal case the whole
phenomenon is magnetic. The inner engine that produces the
relativistic outflow is driven by the Blandford-Znaek mechanism
\cite{BlandfordZnajek} from a magnetized black hole - accretion
disk system \cite{lbw00,lwb00,vp01,LyutikovBlandford03}, or by
 a rapidly rotating highly magnetized pulsar
\cite{Usov92,Thompson94,Katz97}. The relativistic outflow is a
Poynting flux \cite{Usov92,Thompson94,Katz97,LyutikovBlandford03}
and the dissipation arises due to magnetic field recombination or
some other instability. Even with a baryonic outflow  magnetic
fields may play a dominant role in the internal engine as a
possible way to power the relativistic outflow is  by reconnection
of $10^{15}$ G magnetic fields within an accretion disk
surrounding a compact object \cite{NPP92}.

We can observe directly only the emitting regions where the
dissipation and the emission take place. The  inner engine that
accelerates and collimates the relativistic outflow and the
outflow itself are hidden. Both could not be observed directly and
we have only indirect clues on their nature. Our inability to
explore directly the inner engine and the nature of the
relativistic outflow (Poynting flux or baryonic) is quite general.
The same problem arises in modelling relativistic jets seen in
many other astronomical objects. Specifically the relativistic
jets in AGNs and galactic micro quasars are the two most similar
to GRB jets. GRBs might hold the key for the resolution of the
question what are these flow made off?

I begin this short review  with a brief discussion of the fireball
model focusing on issues most directly relevant to magnetic
fields. I refer the reader to \cite{p04} for a more general and
extended review and to \cite{l04} for a rather different
perspective on magnetic fields in GRBs.  I  discuss  the
implication of synchrotron emission on magnetic fields within the
afterglow and the prompt emission. I examine what can we learn
from polarization measurements. After that I turn  to the early
afterglow and discuss the role of the prompt optical flash as a
diagnostic of the nature of the relativistic outflow. I conclude
with a short discussion of future prospects for progress on these
issues.

\section{The internal-external shocks model.}
\label{sec:fireball}

\begin{figure}[h]
  \includegraphics
  [height=0.3\textheight]{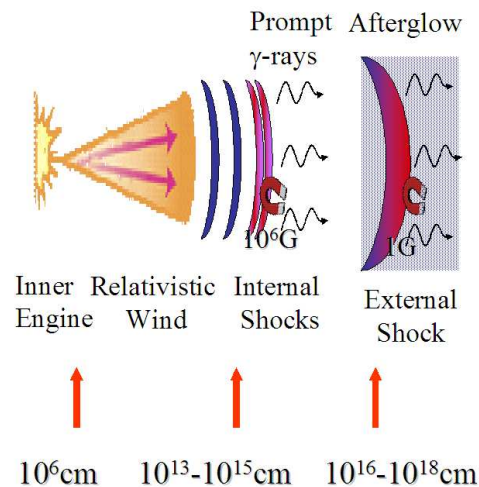}
  \caption{A schematic diagram of the internal-external shocks model. The magnets
  depict the essential magnetic fields in this model and their typical values.}
  \label{fig:1}
\end{figure}

Fig. \ref{fig:1} depicts an overall view of the internal-external
shocks model. A compact source whose size is $\sim 10^{6}$ cm
emits a collimated ultra-relativistic wind with a Lorentz factor
of at least 100. For a long duration GRB this might take place
within a collapsing star, as suggested by the collapsar model
\cite{MacFadyen_W99} and confirmed by the association of long
duration GRBs with type Ic supernovae
\cite{Galama98bw,Bloom99,Stanek03SN,Hjorth03SN}. In this case the
jet has to punch a hole in the stellar envelope, whose radius is
$\sim 10^{10}$ cm. This envelope is not depicted in this picture.

The kinetic energy of the wind is partially dissipated via
internal shocks that take place around $10^{13}-10^{15}$ cm. These
shocks accelerate the electrons to ultra-relativistic energies
(the typical Lorentz factor of an electron is $\sim 1000$). The
electrons emit the observed prompt $\gamma$-rays via synchrotron
radiation. The needed magnetic field is carried from the inner
engine or is generated and amplified by the shocks.

At a distance of $\sim 10^{16}$ cm the loading of the circum-burst
material on the outgoing flow becomes effective and the ejecta
begins to slow down. At first a two shocks system forms. A forward
shock propagating into the circum-burst matter,  a reverse shock
propagating into the ejecta and  a contact discontinuity between
the two. This is a short lived phase during which the early
afterglow arises.

After the reverse shock crosses the ejecta a rarefaction wave
passes and the ejecta cools down adiabatically. At this stage, at
distances of $10^{16}-10^{18}$ cm from the center, the forward
shock collects more and more circum-burst material. It expands
adiabatically and approaches the self similar Blandford-McKee
solution \cite{BLmc1}. This is the ultra-relativistic analog of
the well known Sedov-Taylor Newtonian blast wave solution. As more
material  accumulates the shock slows down, with the Lorentz
factor decreasing like $R^{-3/2}$ and $t^{-3/8}$.

Sometime during this phase the Lorentz factor $\Gamma$, drops
below $\theta^{-1}$ the opening angle of the jetted outflow. At
this stage we encounter a jet break in the afterglow light curve.
For $\Gamma>\theta^{-1}$ most of the emitted radiation is beamed
within $\theta$ but for $\Gamma<\theta^{-1}$ some of the radiation
is emitted sideways. Additionally for $\Gamma<\theta^{-1}$ the
outflow expands sideways and if the expansion is rapid enough the
additional collected matter causes a faster slowing down.

Eventually, around $10^{18}$ cm the blast wave collects sufficient
material to slow it down so that it reaches the Newtonian
transition. This takes place a few month to a year after the
burst. At this stage practically all the radiation is in the radio
frequencies. Later the blast wave approaches the Taylor-Sedov
solution.

The magnetic field parameters in the different regimes are
estimated from the observed emission, that is assumed to be
synchrotron (see the discussion below). In Table \ref{table:1} I
compare various quantities relating to the magnetic fields in the
various regimes.

\begin{table} [h]
\begin{tabular}{|c|c|c|c|c|c|c|c|c|}
  \hline
                 & R                   & B       & $R_L$    & $\lambda_B$
  & $\Delta$      & $R_L/\Delta$ & $\delta$      & $\Delta/\delta$  \\
  \hline
  Internal Shocks& $10^{13}-10^{15}$cm & $10^6$G & 1 cm     &$10^{-3}$cm
  & $10^{11}$cm & $10^{-7}$    & 100 cm   & $10^9$ \\
  \hline
  Afterglow      & $10^{16}-10^{18}$cm & 1G      & $10^6$cm &$10^2$cm
  & $10^{16}$cm & $10^{-9}$    & $10^6$cm & $10^9$ \\
  \hline
\end{tabular}
\caption{Magnetic parameters within GRBs: The size of the region
$R$, the magnetic field $B$, the electron's Larmour radius,
$R_L\equiv \gamma m_e c^2 /e B$, $\lambda_B$ the maximal
correlation length needed for jitter radiation, the skin depth
$\delta\equiv c/\omega_p$, and the width of the emitting region
(in the observer frame) $\Delta$.} \label{table:1}
\end{table}

\begin{figure}
  \includegraphics
  [height=0.3\textheight]{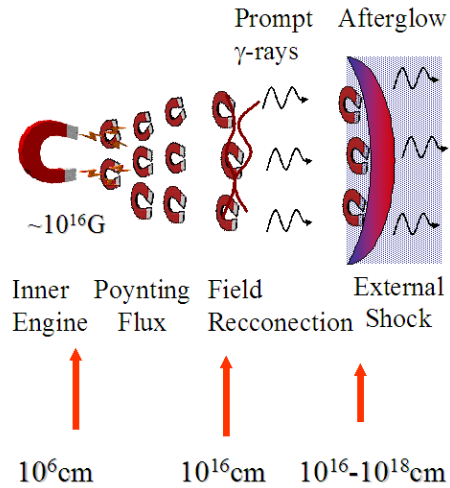}
  \caption{A schematic diagram of the Poynting flux model.}
  \label{fig:2}
\end{figure}

An alternative model is based on a Poynting flux outflow. In this
case the inner engine is magnetic and the dissipation is in the
form of magnetic field recombination. General arguments suggest
that the sizes of the inner engine and of the different emitting
regimes should be comparable to those within the baryonic flow
model. The magnetic fields are however, larger as they not only
contribute to the synchrotron emission they also carry most of the
energy.  Fig. \ref{fig:2} depicts a schematic description of this
model. It is also possible of course that the Poynting flux energy
and the baryonic energy are comparable leading to a mixed model
\cite{SpruitDaigneDrenkhahn01}.

\section{Synchrotron Emission in GRBs}

\begin{figure}
  \includegraphics
  [height=0.35\textheight]
  {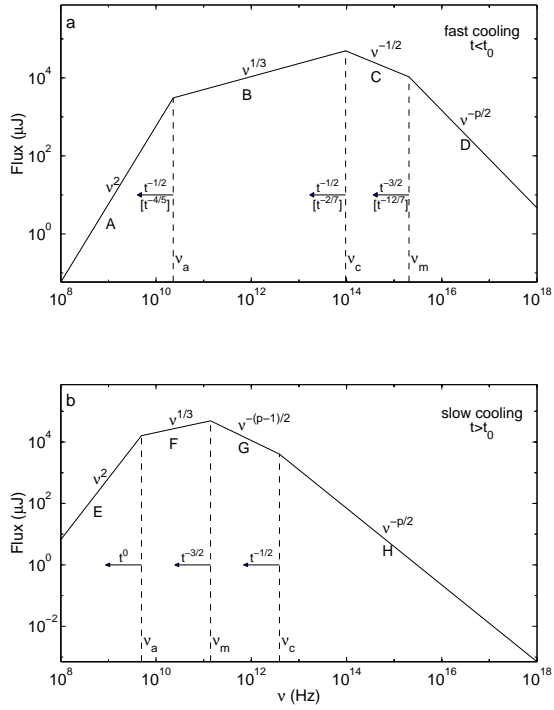}
  \caption{A schematic synchrotron spectrum during the fast (top) and slow
  (bottom) cooling phases (from \cite{SPN98}).}
 \label{fig:3}
\end{figure}

The shock accelerated electrons emit, via synchrotron emission,
the observed radiation. Fig. \ref{fig:3} (from \cite{SPN98})
depicts the typical synchrotron spectrum for fast cooling, which
is applicable during the GRB prompt phase and for slow cooling
that is applicable for the afterglow. It is generally assumed that
the energy densities of the relativistic electrons and of the
magnetic field can be characterized by equipartition parameters,
$\epsilon_e$ and $\epsilon_B$ that measure the ratios of these
energies to the total energy. It is also assumed that  the
electrons energy distributed behaves like a power law with an
index $-p$. Fits to the observed afterglow spectra suggest that
$\epsilon_e \sim 0.2$, $\epsilon_B \sim 0.001$ and $ p \sim
2.3-2.5 $ \cite{PanaitescuK01,yhsf03}.

Schematically, the spectrum is composed of four segments separated
by three typical frequencies. The cooling frequency that is the
synchrotron frequency of an electron that cools on a hydrodynamic
time scale, the synchrotron frequency of the typical electron and
the self absorption frequency (see \cite{SPN98} and
\cite{GranotSari02} for details).

\subsection{Magnetic Fields in the Afterglow}

 The Blandford-McKee ultra-relativistic self-similar
 solution combined with the synchrotron
 radiation model provides an excellent description of the
 afterglow light curves and spectra. As one can see from Table
 \ref{table:1} the magnetic fields in this region are relatively
 large. The value of 1 G is much larger than
the one obtained by  a simple shock compression of the
intergalactic magnetic field. Even if the burst takes place within
a wind that was ejected from the star prior to its explosion the
magnetic fields carried out by the wind at $10^{16}$cm would be
too weak. An immediate question that arises is what is the origin
of this strong magnetic field. This is particularly puzzling as
the afterglow is emitted by  shocked circum-burst matter whose
magnetic field cannot be carried from the inner engine (as can
take place in the internal shocks). The field has to be generated
in site by the shock.

Weibel \cite{w59} suggested already in 1959, that an instability
leading to the growth of the magnetic field would arise whenever
two flows cross each other. Kazimura et al., \cite{ksnb98} have
shown that such an instability could lead to growth of the
magnetic field and to particle acceleration. As the  collisionless
shocks that arise in GRB afterglows are composed of two flows that
cross each other Medvedev and Loeb \cite{MedvedevLoeb99} suggested
that the Weibel instability is the origin of the strong magnetic
fields inferred in GRB afterglows. However, Gruzinov \cite{g01}
raised the concern that this magnetic field will maintain its
equipartition magnitude only over a skin depth $\delta$ which for
GRB afterglows is around $10^6$ cm -  ten orders of magnitude
smaller than the width of the emitting region (see table
\ref{table:1}. While I am not aware of a simple analytic
explanation of this puzzle recent numerical simulations
\cite{nhr+03,jlt04,fhhn4,nffsm05} suggest that equipartition size
magnetic field remains over a region of thousands of skin depth.
Moreover, the same instability accelerates particles to
relativistic energies \cite{nhr+03,hhfn04}, providing another
ingredient of the synchrotron emission. While current simulations
cannot imitate the realistic conditions they clearly suggest that
the required magnetic fields can be amplified in afterglow shocks
and hence in other relativistic collisionless shocks.

\subsection{Synchrotron Prompt Emission}

The energy density within the internal shocks region is much
larger than the energy density within the afterglow, and so are
the corresponding magnetic fields that can be as large as
$10^6-10^7$G (depending on the exact location of the shocks and on
their equipartition parameters). Here the field can be generated
at the shocks (by the same Wiebel instability) or it can be
carried out from the inner engine. The toroidal component of the
field decays only as $R^{-1}$ and hence a field of $10^{14}$ G at
$10^6$ cm can easily reach  $10^{6}$ G at $10^{14}$ cm. Note that
such a magnetic  field is energetically subdominant and it does
not require that the flow will be Poynting flux. A Poynting flux
dominated flow requires at least $10^{15}$ G at the inner engine
and at least $10^7$ G at $10^{14}$ cm.

While  synchrotron emission fits nicely the observed afterglow
spectra and light curves it is not clear that it fits the observed
prompt $\gamma$-ray spectrum. Specifically, the low energy part of
the synchrotron spectrum behaves like $\nu^{1/3}$ (see Fig.
\ref{fig:3}) \cite{Katz94s,CohenKatzP98}. However many GRBs show a
steeper low energy spectrum. That is many spectra fall below the
``synchrotron line of death''
\cite{PreeceEtal98,PreeceEtal00,Preece02}. The constraint is even
more severe, considering the fact that for the required high
efficiency the system must be fast cooling  and this requires a
spectrum less steep than $\nu^{-1/2}$ \cite{CohenKatzP98}. These
observations have led some authors
\cite{Ghisellini_Celotti99,sp04} to rule out synchrotron as the
source of the prompt $\gamma$-ray emission. However, the low
energy spectral observations of the more sensitive (in this energy
band) HETE \cite{BarraudEtal03} do not show numerous bursts below
the ``lines of death''. We may have to wait for better data to
determine this issue.

\begin{figure}[h]
  \includegraphics
  [height=0.25\textheight]{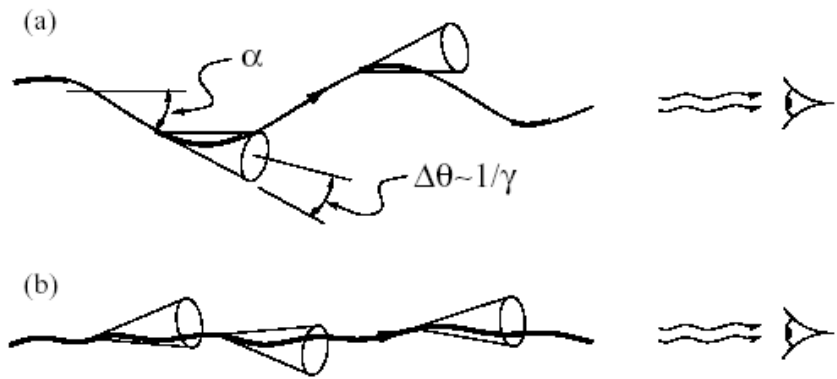}
  \caption{A schematic diagram of the motion of
  an electron in a random magnetic field and Jitter radiation (a) Emission from
  selected parts of the trajectory are seen by the observer. (b) Emission from the
  the entire trajectory is seen by the observer. From
  \cite{Medvedev00}}
  \label{fig:4}
\end{figure}

The problems that the synchrotron model encountered in fitting the
observed GRB spectra have led to the suggestions that
Synchrotron-self-Compton  \cite{gc99,sp04} and inverse-Compton of
an external photon field \cite{Shemi94,ShavivDar95,Lazzatietal03}
are the emission processes. Medvedev \cite{Medvedev00}  proposed
that the prompt emission is produced by Jitter radiation - the
analogue of synchrotron radiation in a random field \cite{nt79}.
Jitter radiation arises in cases that the magnetic field's
correlation length  $\lambda_B $ is smaller than the region over
which an electron emits the synchrotron radiation seen by a given
observer, $R_L/\gamma$ (see Fig. \ref{fig:4}).   The spectra of
Jitter radiation is more complicated than the synchrotron spectra
and Medvedev suggests that it could have a low energy spectrum
that is steeper than $\nu^{-1/2}$ or even than $\nu^{1/3}$. See
however \cite{f05} for remarks on Medvedev's results.

\section{Polarization}
\subsection{Afterglow Polarization }

As synchrotron emission is intrinsically polarized with a suitable
geometry it would lead to polarized afterglow
\cite{GruzinovWaxman99,Sari99,GhiselliniLazzati99,Gruzinov99}. In
particular the required geometry could occur when we observe a
relativistic jet.  By now polarized optical emission has been
observed from several afterglows
\cite{CovinoEtal99,WijersEtal99,RolEtal00,Greineretal03,Bersier03}.
Somewhat surprisingly the polarization magnitude and angle do not
follow the simple predictions
\cite{Sari99,GhiselliniLazzati99,Rossietalpolarization02}. While
the predicted polarization was at the level of a ten percent the
observed polarization was only around a few percent. Furthermore,
the jet structure suggests a specific polarization pattern for a
uniform jet \cite{Sari99,GhiselliniLazzati99} and another one for
a universal structured jet \cite{Rossietalpolarization02}. None
was observed so far.  The patchy shell model in which the local
geometry of the jet is dominated by hot spots \cite{KP00b}
provides a possible explanation for the lower polarization level
and the jumps in the  polarization angle \cite{NakarOren03}.

Granot and Taylor \cite{gt04} reported recently  upper limits on
the polarization of GRB radio flares (the early part of the radio
afterglow).  Their best limit, for GRB 991216 is around 10\%. They
point out that this upper limit already rules out a universal
structure jet model with a homogenous magnetic field. More refines
observations will enable us to distinguish between different jet
models.

\subsection{Polarized Prompt Emission}

Coburn and Boggs \cite{CoburnBoggs03} announced the discovery of
very strong ($\sim 80$\%) polarization from the prompt emission of
GRB 021206. It has been argued that this level of polarization
shows that the magnetic filed must be uniform
\cite{CoburnBoggs03,Granot03} and that this further implies that
the flow must be  Poynting flux dominated \cite{LyutikovPB03}.
However, two independent groups
\cite{Rutledge03Polarization,wha+pol04} reanalyzed the same data
and found no statistical viable indication of polarization.
Additionally, synchrotron emission from a relativistic jet
\cite{NakarPiranWaxman03} can produce a $\sim 50\%$ polarized
light even with a random magnetic field (provided that the field
is within the plane of the shock). This is only slightly lower
than the maximal emission from a homogenous magnetic field
configuration \cite{Granot03,NakarPiranWaxman03} which is around
60\%. Furthermore, a homogenous (toroidal) field configuration
would arise in any case when the field is dragged from the source
\cite{SpruitDaigneDrenkhahn01}. Thus, homogeneity does not
necessarily imply that the flow is Poynting flux dominated. We
still need a better way to distinguish between Poynting flux and
baryonic outflow.

\section{The Early Afterglow - a Diagnostic Tool or the Nature of the Outflow}

One of the most interesting  and most vividly debated questions
conceding GRBs  is the nature of the relativistic outflow -
Poynting flux or a baryonic outflow. One of the best way to
explore this question is by observation and analysis of the early
afterglow \cite{np04,np05,np05a}.

\begin{figure}
  \includegraphics
  [height=0.3\textheight]{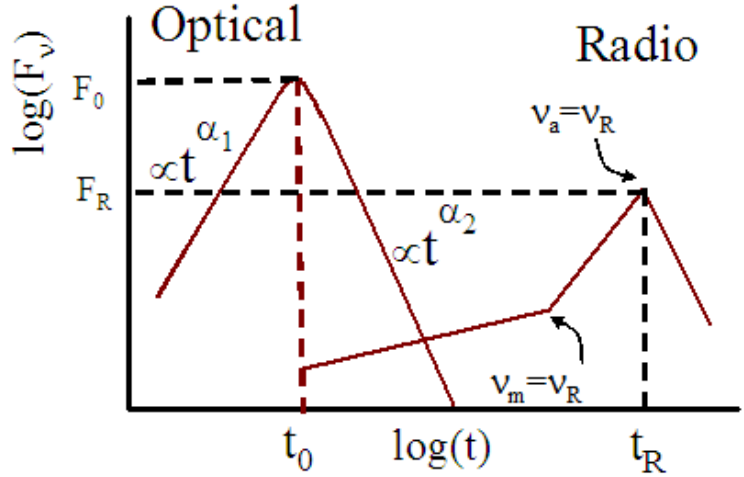}
  \caption{A schematic diagram of the early afterglow optical and Radio light curves.}
  \label{fig:5}
\end{figure}

The initial interaction of the relativistic outflow with the
circum-burst material results in two shocks. A short lived reverse
shock that propagates into the outflow and a long lived forward
shock  that propagates into the circum-burst material. The later
eventually turns into the relativistic blast wave that produces
the afterglow. Recently Nakar and Piran \cite{np04} presented a
new analysis of the reverse shock emission (see also \cite{zk04}).
They show that a baryonic jet has a unique signature  and that a
combined analysis of the optical and the radio data could,
therefore, enable us to explore the nature of the outflow.

The reverse shock optical emission peaks at $t_o$ when the reverse
shock reaches the inner edge of the outflow.  The first unique
signature of the reverse shock is  a characteristic decay of the
optical flux after $t_o$ as $\sim t^{-2}$
\cite{SP99b,KobayashiSari01}. This decay arises due to the
adiabatic cooling of the shocked material. The behavior is quite
different at the radio. Initially, the typical synchrotron
frequency, $\nu_m$ is above the observed radio frequency $\nu_R$,
and below the self absorption frequency, $\nu_a$. As the shocked
material cools $\nu_m$ decreases and the radio emission increases
as $t^{1/2}$ until $\nu_m  = \nu_R$. At this stage the radio flux
increases like $t^{3/2}$ until, at $t_R$ $\nu_a=\nu_R$ and from
that moment on the radio emission decreases like $t^{-2}$ just
like the optical. This behavior is schematically depicted in Fig.
\ref{fig:5}

For baryonic outflow there is a unique relation between the time
and flux of the optical peak and the time and flux of the radio
peak, $t_o$, $F_{o}(t_0)$, $t_R$ and $F_R(t_R)$: $(F_R/F_o)
(t_R/t_o)^{(p-1)/2+1.3}
 = C (\nu_o/\nu_R)^{(p-1)/2} \approx 1000$,
where $p$ is the index of the electron energy distribution and $C$
is a constant of order unity. Current uncertainties put $C$
between 0.5 and 2. By now the only burst with well determined
optical early afterglow is GRB 990123. Its optical flash decayed
as  $t^{-2}$. The corresponding factor that relates the optical
and radio peak fluxes and peak times for GRB 990123 is 1300 very
close to the expected value (1000-2000). Both results indicate a
baryonic outflow in this particular burst.

\section{Some Future Prospects}

Magnetic fields are clearly a key ingredient of GRBs. They are
essential in the emitting regions and it may very well be that
they appear in all parts of the GRB phenomenon, including  the
central engine and the relativistic outflow.

As Nakar \& Piran have shown  \cite{np04,np05,np05a} combined
measurements of the optical afterglow and the radio flare could
enable us to determine the early afterglow parameters for bursts
in the nearby future. Swift and several quick response ground
telescopes are ideally suited for these optical  observations and
hopefully radio telescopes will be allocated sufficient time and
quickly enough to detect the corresponding radio flares. This
would hopefully shed some light on the central question - the
nature of the relativistic outflow - baryonic or Poynting flux.

Observations of prompt polarization are quite unlikely as one need
an extremely bright burst, and such bursts are very rare. On the
other hand the quick notification of Swift may allow measurements
of many optical polarizations and possibly radio flare
polarizations. These could shed further light on the signature of
magnetic fields and on the angular structure of GRB jets.

As for the formation of magnetic fields in collisionless shocks,
the success of the synchrotron model in explaining the basic
properties of GRB afterglow demonstrates that nature knows how to
generate magnetic fields in collisionless shock, even if we don't
understand yet how. The progress made in numerical simulations
suggest that we are on the right track and indeed it is the Weibel
instability that is responsible for this field amplification. This
might have some implications for seeding magnetic fields elsewhere
in the Universe


\begin{theacknowledgments}
I thank Ehud Nakar for many helpful discussions. This research was
supported in part by a US-Israel BSF grant.
\end{theacknowledgments}




\end{document}